\documentclass[12pt,jcp,showpacs]{revtex4}

\newcommand{\PR}[3]{Phys. Rev. \textbf{#1}, {#2} (#3).}
\newcommand{\JPhC}[3]{J. Phys. Chem. \textbf{#1}, {#2} (#3).}
\newcommand{\JPCA}[3]{J. Phys. Chem. A \textbf{#1}, {#2} (#3).}
\newcommand{\JCC}[3]{J. Comput. Chem. \textbf{#1}, {#2} (#3).}
\newcommand{\JCP}[3]{J. Chem. Phys. \textbf{#1}, {#2} (#3).}
\newcommand{\Carbon}[3]{Carbon \textbf{#1}, {#2} (#3).}
\newcommand{\ARAA}[3]{An. Rev. Astr. Ap. \textbf{#1}, {#2} (#3).}
\newcommand{\AandA}[3]{A\&A \textbf{#1}, {#2} (#3).}
\newcommand{\CP}[3]{Chem. Phys. \textbf{#1}, {#2} (#3).}
\newcommand{\CPL}[3]{Chem. Phys. Lett. \textbf{#1}, {#2} (#3).}
\newcommand{\Astrophys}[3]{ ApJ \textbf{#1}, {#2} (#3).}
\newcommand{\ChemPhysChem}[3]{ChemPhysChem \textbf{#1}, {#2} (#3).}
\newcommand{\TheorChem}[3]{Theor Chem Acc \textbf{#1}, {#2} (#3).}
\newcommand{\JACS}[3]{J. Am. Chem. Soc. \textbf{#1}, {#2} (#3).}
\newcommand{\ASPC}[3]{ASP Conference series \textbf{#1}, {#2} (#3).}

\newcommand{\CE}[3]{Chem. Ed.: Research and Practice in Europe \textbf{#1}, {#2} (#3).}

\newcommand{\lf}[2]{#2. #1,}
\newcommand{\lff}[3]{#2. #3. #1,}

\usepackage{graphicx}
\begin{document}

\title{T-shaped dimer of coronene}

\author{O. I. Obolensky}
\email{o.obolensky@fias.uni-frankfurt.de}
\altaffiliation[On leave from: ]{A.F. Ioffe Institute, Politechnicheskaja str. 26,
St. Petersburg 194021, Russia}
\affiliation{Frankfurt Institute for Advanced Studies, Johann Wolfgang Goethe-University,
Max von Laue str. 1, D-60438 Frankfurt am Main, Germany}

\author{V. V. Semenikhina}
\affiliation{Frankfurt Institute for Advanced Studies, Johann Wolfgang Goethe-University,
Max von Laue str. 1, D-60438 Frankfurt am Main, Germany}

\author{A. V. Solov'yov}
\email{solovyov@fias.uni-frankfurt.de}
\altaffiliation[On leave from: ]{A.F. Ioffe Institute, Politechnicheskaja str. 26,
St. Petersburg 194021, Russia}
\affiliation{Frankfurt Institute for Advanced Studies, Johann Wolfgang Goethe-University,
Max von Laue str. 1, D-60438 Frankfurt am Main, Germany}

\author{W. Greiner}
\affiliation{Frankfurt Institute for Advanced Studies, Johann Wolfgang Goethe-University,
Max von Laue str. 1, D-60438 Frankfurt am Main, Germany}

\begin{abstract}
An evidence
of importance of the T-shaped configuration
of coronene dimer is presented.
That is, the dimer's lowest energy configuration
is not necessarily a stack, as it might had been expected \textit{a priori}. 
This is a surprising result for dimer of such a large
polycyclic aromatic hydrocarbon (PAH) as coronene (C$_{24}$H$_{12}$).
The energy of the T-shaped configuration at all
considered levels of density functional theory
(B3LYP,PBE/6-31+G(d),D95,cc-pVDZ,cc-pVTZ)
was systematically lower than the energies of three plausible
stack configurations.
In order to get a better description of
the van der Waals interaction,
the
density functional theory (DFT)
results were adjusted by adding a phenomenological Lennard-Jones-type
term into the total energy of the system.
However,
the van der Waals correction is somewhat arbitrary
and its magnitude can not be rigorously justified.
Depending on the choice of the parameters in the phenomenological term
both the T-shaped and the parallel-displaced (PD) stack configurations can be
the global energy minimum of the system.
A simple model is proposed which is useful for
qualitative understanding of possible geometries of the coronene dimer
and larger coronene clusters. The model represents coronene
dimer as two sets of charged rings interacting
via Coulomb and Lennard-Jones potentials.
The model provides an intuitively clear explanation why
the T-shaped dimers can be of importance
even for some of moderately large PAHs such as coronene and, may be,
for circumcoronene.
The unexpectedly strong binding of the coronene dimer
in the T-shaped configuration 
is connected to the round shape of the
coronene molecules.
Indeed, rotation or parallel-displacement of the non-round monomers
results in significantly smaller Coulomb repulsion
as compared to the "face to face" sandwich
configuration.
On the other hand, rotation and/or displacement are much less
effective for the stacked round molecules.
Therefore, the round shape of the coronenes
leads to an increased role of the electrostatic repulsion
in the stack configurations
and the T-shaped configuration can become competitive.
The proposed model can be
easily generalized to other polycyclic
aromatic hydrocarbons, DNA bases, etc.
\end{abstract}


\maketitle

\section{Introduction}

Molecules of polycyclic aromatic hydrocarbons and their ions
are of current interest as they are believed to be
one of the most probable species in the interstellar space
responsible for
converting the UV radiation of stars into
the intensive IR radiation,
known as unidentified infrared bands
\cite{Duley81,Puget84,Allamandola85,Langhoff96,%
Allamandola99,Allamandola03,Duley04}.
The physical mechanisms and effectiveness of energy transfer
from the UV to the IR radiation occuring via transient heating \cite{Sellgren84}
are currently being studied,
see, e.g., \cite{Langhoff96,Li04,Allamandola05,Rapacioli05}.

Astrophysical observations suggest that the interstellar medium
contains a variety of small
carbonaceous
particles with the sizes ranging
from dust grains down to single molecules \cite{Puget89}.
The most abundant particles
contain
between 100 and 1000 carbon atoms \cite{Puget98} in the forms of
collections (clusters) of PAH-like molecules \cite{Rapacioli05} of various degrees of
hydrogenation \cite{Duley04}.
Hence, studying
properties of PAH clusters presents a necessary step
in further research of interstellar medium.

The current knowledge of structure and ways of formation of PAH clusters
is very limited.
Laboratory studies have just started to appear
and only anthracene \cite{Piuzzi02,Kaya04}, naphthalene \cite{Kaya04}
and coronene \cite{Kaya99,Brechignac05} clusters
have been produced so far, so that
the array of available experimental data is sparse. 
Theoretical investigations of oligomers of the larger PAHs are also scarce and,
furthermore, their results are very sensitive to a subtle interplay
between the van der Waals polarization forces (also sometimes called
dispersive interaction) and the electrostatic interaction.
Even for benzene, the simplest aromatic hydrocarbon,
an unequivocal choice between the T-shaped
and the parallel-displaced stack configurations can not be made
since the calculated binding energies for these configurations
strongly depend on the choice of the method of calculation and
on the basis set and they do not seem to be converging
even at the most advanced levels of theory
\cite{Jaffe96,Hobza96,Sherrill02}.
Alternative approaches based
on mixing of the \textit{ab initio} and the phenomenological treatments
have been suggested which allow a more flexible tuning of
interactions (see, e.g., \cite{Lim03,Grimme04,Shishkin05,Grimme05,Wales05}),
but their universality and predictive power
are not quite clear yet.

We have performed
density functional theory calculations
of energies for four plausible configurations \cite{Wales05} of the coronene dimer.
These included three stack configurations (superimposed stack - "sandwich" (SS),
twisted stack (TS) and parallel-displaced (PD) stack) and the
T-shaped configuration.
To our knowledge, this is the first comparative study
of the configurations of the coronine dimer within the \textit{ab initio} approach.
The only other \textit{ab initio} calculation of coronene dimer parameters was done
for the PD stack configuration \cite{Grimme04}.

The energy of the T-shaped configuration at all
considered levels of theory
(combinations of the B3LYP and PBE functionals 
with 6-31+G(d), D95, cc-pVDZ, and cc-pVTZ basis sets)
was systematically lower than the energies of the 
stack configurations.
In order to get a better description of
the van der Waals interaction,
the DFT results
were adjusted by adding a phenomenological Lennard-Jones-type
term into the total energy of the system.
The problem of this approach is that the van der Waals correction
is somewhat arbitrary and different authors choose
different parameters even for the same systems (cf., e.g., \cite{Grimme04}
and \cite{Wales05}). Depending on magnitude of the van der Waals term,
both the T-shaped and the stack configurations can be
the global minimum configuration of the system.
Even though the definitive conclusion can not be drawn at the moment,
it is quite surprising on its own that the T-shaped configuration
is energetically competitive for such a large system
for which a graphite-like stacked structure was expected \cite{Grimme04}.

We propose a simple model useful for qualitative understanding
of dimerization and clusterization of polycyclic aromatic hydrocarbons.
The model represents a PAH molecule
as a set of circular or deformed rings
corresponding to the chains of carbon atoms surrounded by a chain
of hydrogens.
For example, the coronene molecule is represented
as a set of four nested circular rings with the outer
ring corresponding to the hydrogens.
The electron density in PAHs
is shifted from the hydrogens towards the interior of the molecules.
In a crude approximation this charge distribution can be
described as a polarized band (with a positively charged outer edge and
a negatively charged inner edge) located
along the perimeter of a molecule.
This charge distribution can also be considered as a dipole
distributed along the border of a molecule.
Two PAH molecules interact electrostatically and via
the van der Waals forces between the sets of the rings
substituting the monomers.

The model
provides a qualitative explanation why
even for such a large PAH
as coronene (C$_{24}$H$_{12}$) the T-shaped configuration
is comparable in energy with the stack configuration.
This happens due to the round shape of the coronene molecule
which leads to an increased electrostatic repulsion
in the stack configurations. For non-round molecules
in the stack configurations, the electrostatic interaction
can be made less repulsive (or even attractive) by rotating
or by parallel displacing the molecules in the dimer,
while for stacked round molecules the rotation or
displacement are much less
effective and the T-shaped configuration can become competitive.

\section{Density functional theory treatment of coronene dimer}

We have performed density functional theory calculations
of binding energies for four
configurations of coronene dimer. The dimer binding energy 
is defined as follows:

\begin{equation}
E_{\rm b}= 2E_{\rm mono} -E_{\rm dimer},
\label{Ebinding}
\end{equation}
\noindent where  $E_{\rm dimer}$ is the total energy of the dimer
and $E_{\rm mono}$ is the energy of the single molecule.
Note that according to this definition the binding energies
of the stable bound states of the
dimer are positive.

\begin{figure}[htb]
\begin{center}
\includegraphics[scale=0.45,clip]{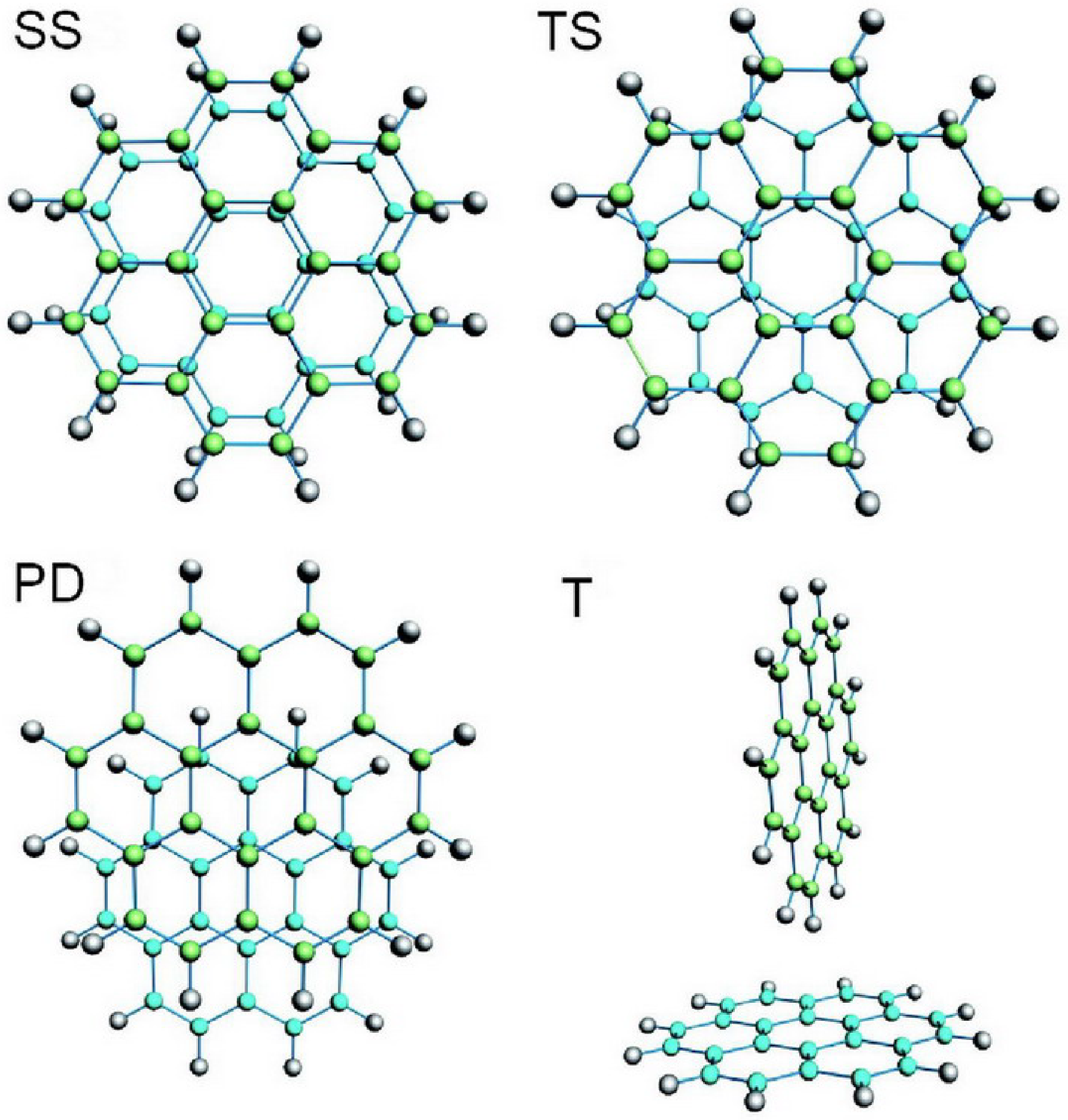}
\end{center}
\caption{(Color online) Geometries of four configurations of the coronene dimer.
In the sandwich (SS) configuration the distance $R$ between the centers
of mass of the monomers equals 4.27~\AA.
In the twisted sandwich (TS) configuration $R=4.21$~\AA \ and the rotation angle
is $30^o$.
In the parallel-displaced (PD) configuration $R=3.95$~\AA \ and the shift
is 2.70~\AA.
In the T-shaped (T) configuration
$R=8.02$~\AA.
The binding energies, calculated at various levels of
theory, can be found in Table \protect{\ref{binding}}.
}
\label{geometries}
\end{figure}

The studied geometries are shown in Figure \ref{geometries} and the
binding energies are summarized in Table \ref{binding}. The
calculations have been done with the B3LYP and PBE functionals
by expanding the molecular
orbitals into the standard 6-31+G(d), D95, cc-pVDZ, and cc-pVTZ
basis sets as implemented in \textit{Gaussian}03
package \cite{Gaussian03,GaussianBook}.

\begin{table}[htb]
\caption{\label{binding} The binding energies (\protect{\ref{Ebinding}})
of the coronene dimer
(in kcal/mol) calculated at various levels of theory.
For notations and geometries of the dimer configurations
see Figure \protect{\ref{geometries}}.
}

\begin{tabular}{|c|c|c|c|c|}
\hline
            &\multicolumn{1}{|c|}{\ \ \ \ \  SS \ \ \ \ \ \  }
            &\multicolumn{1}{|c|}{\ \ \ \ \  TS \ \ \ \ \ \ }
            &\multicolumn{1}{|c|}{\ \ \ \ \  PD \ \ \ \ \ \ }
            &\multicolumn{1}{|c|}{\ \ \ \ \  T \ \ \ \ \ \ }\\
        \hline
B3LYP/6-31+G(d) & \multicolumn{1}{|c|}{-12.55}
        &\multicolumn{1}{|c|}{-9.16}
        &\multicolumn{1}{|c|}{-9.41}
        &\multicolumn{1}{|c|}{0.33} \\
        \hline
B3LYP/D95 & \multicolumn{1}{|c|}{-2.99}
        &\multicolumn{1}{|c|}{-2.64}
        &\multicolumn{1}{|c|}{-1.91}
        &\multicolumn{1}{|c|}{0.43} \\
        \hline
B3LYP/cc-pVDZ & \multicolumn{1}{|c|}{-4.49}
        &\multicolumn{1}{|c|}{-2.72}
        &\multicolumn{1}{|c|}{-2.55}
        &\multicolumn{1}{|c|}{0.43} \\
        \hline
PBE/6-31+G(d)  & \multicolumn{1}{|c|}{-5.34}
        &\multicolumn{1}{|c|}{-0.91}
        &\multicolumn{1}{|c|}{1.7}
        &\multicolumn{1}{|c|}{1.98}\\
        \hline

PBE/D95 & \multicolumn{1}{|c|}{0.88}
        &\multicolumn{1}{|c|}{1.39}
        &\multicolumn{1}{|c|}{2.32}
        &\multicolumn{1}{|c|}{2.83}\\
        \hline
PBE/cc-pVDZ & \multicolumn{1}{|c|}{0.12}
        &\multicolumn{1}{|c|}{0.66}
        &\multicolumn{1}{|c|}{1.72}
        &\multicolumn{1}{|c|}{1.79}\\
        \hline
PBE/cc-pVTZ & \multicolumn{1}{|c|}{-0.51}
        &\multicolumn{1}{|c|}{0.76}
        &\multicolumn{1}{|c|}{1.22}
        &\multicolumn{1}{|c|}{1.48}\\
        \hline
PBE/TZV(2d,2p) from Ref. \cite{Grimme04} & \multicolumn{1}{|c|}{}
        &\multicolumn{1}{|c|}{}
        &\multicolumn{1}{|c|}{-4.89}
        &\multicolumn{1}{|c|}{}\\
        \hline
BLYP/TZV(2d,2p) from Ref. \cite{Grimme04} & \multicolumn{1}{|c|}{}
        &\multicolumn{1}{|c|}{}
        &\multicolumn{1}{|c|}{-17.52}
        &\multicolumn{1}{|c|}{}\\
        \hline
\end{tabular}
\end{table}

The energy of the T-shaped configuration at all
considered levels of theory
was systematically lower than the energies of the three 
stack configurations.
It is a quite surprising result
for such a large PAH molecule.
For the smallest aromatic hydrocarbon, benzene,
the T-shaped configuration is almost isoenergetic with
the parallel-displaced stack configuration \cite{Jaffe96,Hobza96,Sherrill02}.
It is commonly believed that increase in number of aromatic
rings favors stack configurations over the T-shaped ones
\cite{Jorgensen90,Lim03,Wales05}.
In accord with this assumption, the T-shaped (or, more
generally, "edge to plane") dimers of pyridine, naphtalene, azulene and anthracene
have been found less binding than the stacked configurations \cite{Piuzzi02,Lim03,Grimme05,Grimme_JACS05}.
In the only other \textit{ab initio} calculation of a coronene
dimer \cite{Grimme04}
a graphite-like (parallel-displaced) stacked structure was
considered only.

To our knowledge, this paper is the first comparative study
of the possible configurations of the coronine dimer performed
within the \textit{ab initio} approach.
There were also studies of the coronene dimer which used phenomenological
approaches \cite{Smyth_JPhC84,Marzec_Carbon00,Wales05}.
In the most recent study of this type \cite{Wales05},
the dimers and clusters of PAH molecules
were modeled within a rigid-body approximation with the
use of explicit
interatomic potentials
(a sum of an electrostatic and a van der Waals terms).
None of these semiempirical studies found
the T-shaped form of the coronene dimer to be energetically
preferable.

It is well known that the pure DFT methods does not
reproduce the attractive polarization interactions well enough,
for a discussion see, e.g., Ref. \cite{Grimme04} and references
therein.
This may present a serious problem in such intricate cases
as determining the binding energy for loosely bound dimers
of PAH molecules.
In order to
estimate the influence of
the polarization van der Waals interaction on the preferred dimer
geometry,
we have included
a phenomenological Lennard-Jones-type
term into the total energy of the system, similar to the procedure
developed in \cite{Grimme04}:
\begin{equation}
E_{\rm{}}=E_{\rm{DFT}} + \sum_{i,j}
\left(\frac{C_{12}^{\,ij}}{r_{ij}^{12}}-\frac{C_6^{\,ij}}{r_{ij}^{\,6}}\right).
\label{Etot}
\end{equation}

\noindent Here $E_{\rm{DFT}}$ is the total energy of the system
calculated within the frame of the DFT.
Index $i$ enumerates atoms in one monomer,
index $j$ enumerates atoms in the other monomer;
correspondingly, $r_{ij}$ is the distance between atom "$i$"
in the first monomer and atom "$j$" in the second monomer.
The coefficients $C_6^{\,ij}$ and $C_{12}^{\,ij}$ can be
tabulated directly (as is in Ref. \cite{vdWaal_JCP83})
or found with the help of a combining rule (as is in Ref. \cite{Grimme04})
\begin{equation}
C_M^{\,ij} = 2 \frac{C_M^{\,i} C_M^{\,j}}{C_M^{\,i} + C_M^{\,j}}, \qquad M=6,12.
\end{equation}
\noindent Unfortunately, there are no fixed values for the coefficients $C_6$ and $C_{12}$
which would be universally applicable in a wide scope of situations.
Even for the same systems different authors choose
different parameters  (cf., e.g., \cite{Grimme04}
and \cite{Wales05}).
The arbitrariness of the van der Waals correction seems to be
a general problem of all such hybrid approaches.

In order to illustrate how sensitive the overall conclusions
may be with respect to the values of the phenomenological parameters
we present in Table \ref{DFTvdW} the binding energies
for each of the considered dimer configurations
for various reasonable choices of the parameters $C_6^{\rm C}$,
$C_6^{\rm H}$, $C_{12}^{\rm C}$, and $C_{12}^{\rm H}$.
Depending on magnitude of the van der Waals term,
both the T-shaped and the stack configurations can be
the global minimum configuration of the system.
Thus, we suppose that a definitive conclusion
can not be drawn at the moment.
An experimental measurement of certain characteristics (such as
frequencies of transitions between rotational levels)
would be useful for determining the global energy minimum
configuration of the dimer.
To facilitate 
identification of the dimer configurations, in Table \ref{rotconst}
the rotational constants for each configuration are given. The
rotational constants $B_{\rm \eta}$, connected to the principal values
of the inertia tensor $\cal{I}$ as $B_{\rm \eta}= {\hbar}/(4\pi
\cal{I}_{\rm \eta\eta})$ ($\eta=x,y,z$), define the rotational levels
$E_{\eta}=B_{\eta} J(J+1)$ corresponding to different rotational quantum numbers $J$.

\squeezetable
\begin{table}[htb]
\caption{\label{DFTvdW}
The binding energies
(in kcal/mol)
for four
configurations of the coronene dimer (see Figure \protect{\ref{geometries}})
calculated according to expressions (\protect{\ref{Ebinding}}) and (\protect{\ref{Etot}})
for several sets of the coefficients $C_6$ (in J/mol nm$^6$) and
$C_{12}$ (in mJ/mol nm$^{12}$). The energies of the van der Waals
terms are given in parentheses. The optimization of the dimer
geometries and calculations of the DFT binding energies have been
done with the PBE/D95 method.}

\begin{tabular}{|p{5.2cm}|c|c|c|c|}
\hline \ \ \ \ \ \ \ \ \ \ \ \ \ \ \ \ \ \ \ \ \ \ \ \  \ \ \ \ \ \
\ \ \ \ &\multicolumn{1}{|c|}{\ \ \ \ \ \ \ \ SS \ \ \ \ \ \ \ \ \ }
            &\multicolumn{1}{|c|}{\ \ \ \ \ \ \ \ TS \ \ \ \ \ \ \ \ \ }
            &\multicolumn{1}{|c|}{\ \ \ \ \ \ \ \ PD \ \ \ \ \ \ \ \ \ }
            &\multicolumn{1}{|c|}{\ \ \ \ \ \ \ \ T \ \ \ \ \ \ \ \ \ }\\
\hline

 \ $C_6^{\rm C}= 2.77$\ \ $C_{6}^{\rm H}=0.14$\ \ $C_{6}^{\rm
C-H}=0.63$\footnote{From Ref. \cite{vdWaal_JCP83}.}\

 \ $C_{12}^{\rm C}= 4.87$\ \ $C_{12}^{\rm H}=0.09$\ \ $C_{12}^{\rm
C-H}=0.68$
         & \multicolumn{1}{|c|} {25.73 (24.85)}
         &\multicolumn{1}{|c|}{19.07 (17.68)}
         &\multicolumn{1}{|c|}{17.72 (15.40)}
         &\multicolumn{1}{|c|}{7.54 (4.71)} \\
         \hline

\ \ \ \ \ \ \ $C_6^{\rm C}= 1.15$\ \ $C_{6}^{\rm H}=0.11$ \footnote{From Ref. \cite{Grimme04} 
($C_6$ is adjusted by the factor 0.7 from one
of the values (1.65) given in Ref. \cite{Yang_JCP02}).}
\footnote{A dump function has been used instead of $C_{12}$ \protect{\cite{Grimme04}}}
        &\multicolumn{1}{|c|}{17.97 (17.09) }
        &\multicolumn{1}{|c|}{10.12 (8.73)}
        &\multicolumn{1}{|c|}{9.97 (7.65)}
        &\multicolumn{1}{|c|}{5.52 (2.69)} \\
        \hline

\ \ \ \ \ \ \ $C_6^{\rm C}=0.88$\ \ $C_{6}^{\rm H}=0.11$
\footnote{Another possible value (1.26) of the coefficient $C_6^{\rm C}$
from Ref. \cite{Yang_JCP02} multiplied by 0.7 as in Ref.
\cite{Grimme04}}
        & \multicolumn{1}{|c|}{2.34 (1.46)}
        &\multicolumn{1}{|c|}{5.73 (4.34)}
        &\multicolumn{1}{|c|}{6.00 (3.68)}
        &\multicolumn{1}{|c|}{3.51 (0.68)}\\

\ \ \ \ \ \ \ $C_{12}^{\rm C}=4.87$\ \ $C_{12}^{\rm
H}=0.09$\footnote{Coefficients $C_{12}$ are from Ref.
\cite{vdWaal_JCP83}.}
        & \multicolumn{1}{|c|}{}
        &\multicolumn{1}{|c|}{}
        &\multicolumn{1}{|c|}{}
        &\multicolumn{1}{|c|}{}\\

\hline

 \ \ \ \ \ \ \ $C_6=0$ \ \ \ \ \ \ $C_{12}=0$ & \multicolumn{1}{|c|}{0.88 (0.00)}
        &\multicolumn{1}{|c|}{1.39 (0.00)}
        &\multicolumn{1}{|c|}{2.32 (0.00)}
        &\multicolumn{1}{|c|}{2.83 (0.00)}\\
\hline
\end{tabular}
\end{table}

We note that the superimposed stack (sandwich) configuration
probably is
not a minimum on the potential energy surface.
An analysis \cite{Wales05} done with the use of a phenomenological potential
has identified it as a transitional state between the twisted
stack and the parallel-displaced stack configurations.
This conclusion is indirectly supported by the results of
the high-level \textit{ab initio} calculations
which demonstrated that in benzene the superimposed stack
is also a saddle point rather than a true minimum \cite{Hobza96}.

\squeezetable
\begin{table}[htb]
\caption{\label{rotconst}
Rotational constants (in GHz) for different configurations
of the coronene dimer calculated for the geometries
optimized at
B3LYP/D95 and PBE/D95 levels of theory.
For notations and geometries of the dimer configurations
see Figure \protect{\ref{geometries}}.
}

\squeezetable
\begin{tabular}{|c|c|c|c|c|c|c|c|c|c|c|c|c|}
\hline
& \multicolumn{3}{|c|}{SS} & \multicolumn{3}{|c|}{TS}& \multicolumn{3}{|c|}{PD}& \multicolumn{3}{|c|}{T} \\
\hline
B3LYP/D95 & \multicolumn{1}{|c|}{\ 0.085\ }
        &\multicolumn{1}{|c|}{\ 0.085\ }
        &\multicolumn{1}{|c|}{\ 0.082\ }
        &\multicolumn{1}{|c|}{\ 0.087\ }
        &\multicolumn{1}{|c|}{\ 0.087\ }
        &\multicolumn{1}{|c|}{\ 0.082\ }
        &\multicolumn{1}{|c|}{\ 0.123\ }
        &\multicolumn{1}{|c|}{\ 0.060\ }
        &\multicolumn{1}{|c|}{\ 0.048\ }
        &\multicolumn{1}{|c|}{\ 0.110\ }
        &\multicolumn{1}{|c|}{\ 0.041\ }
        &\multicolumn{1}{|c|}{\ 0.036\ } \\
        \hline
PBE/D95& \multicolumn{1}{|c|}{\ 0.087\ }
        &\multicolumn{1}{|c|}{\ 0.087\ }
        &\multicolumn{1}{|c|}{\ 0.082\ }
        &\multicolumn{1}{|c|}{\ 0.089\ }
        &\multicolumn{1}{|c|}{\ 0.089\ }
        &\multicolumn{1}{|c|}{\ 0.082\ }
        &\multicolumn{1}{|c|}{\ 0.113\ }
        &\multicolumn{1}{|c|}{\ 0.074\ }
        &\multicolumn{1}{|c|}{\ 0.059\ }
        &\multicolumn{1}{|c|}{\ 0.111\ }
        &\multicolumn{1}{|c|}{\ 0.041\ }
        &\multicolumn{1}{|c|}{\ 0.037\ } \\
        \hline
\end{tabular}
\end{table}

\section{Rings model of polycyclic aromatic hydrocarbons}

The model 
approaches
sometimes can be very useful in 
revealing the underlying physics of a process and in pointing
to the directions of search.
In some situations, when
the \textit{ab initio} calculations
are not feasible,
the model calculations are the only approach
one has to resort to.
We present here
a model which
elucidates the role of the shapes of PAH molecules in the process
of dimerization
and in formation of bigger PAH clusters.

\begin{figure}[htb]
\begin{center}
\includegraphics[scale=0.35,clip]{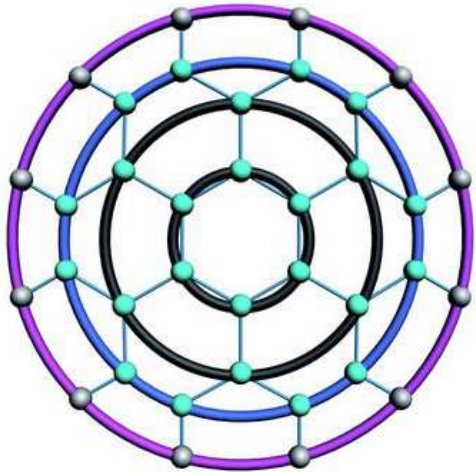}
\end{center}
\caption{(Color online) The round-shaped coronene molecule
represented as a set of four nested circular rings.
The radii of the rings are 1.438, 2.876, 3.595, and 4.691 \AA \
(R$_{\rm C-C}$=1.438\AA, R$_{\rm C-H}$=1.096\AA).
The outermost ring corresponds to the hydrogens and is positively charged
(Q=1.8 $e$).
In the simplest case, this charge is compensated by
the negative charge of the next ring, corresponding to the chain
of outer carbon atoms. Two inner rings, then, are neutral.
}
\label{model}
\end{figure}

The model
represents PAH molecules
as a set of circular or deformed rings which correspond to the chains
of carbon atoms surrounded by a chain
of hydrogens.
For example, the coronene molecule can be described as consisting of
four concentric circular rings representing three chains of carbon atoms and
one chain of hydrogens, see Figure \ref{model}.

The electron density
in PAHs
is shifted from the hydrogens
towards the interior of the molecules.
Therefore, the outer ring
(corresponding to the hydrogens)
is always charged positively.
The inner rings can, in principle, bear any charge, but their net
charge, in the case of a neutral molecule, must be negative and
compensate the positive charge of the outermost ring.

The magnitude of charge for each ring is defined on the basis of the
results of the DFT calculations as the sum of the partial charges of
the atoms belonging to the corresponding chain.
Various methods can be used for assigning the partial charges to the atoms.
For example, Mulliken suggested to calculate partial charges
as the integral over the electron density contained
within a sphere of a given radius, specific to the corresponding type
of atoms \cite{Mulliken_JCP55}. Alternatively, one can
expand the electron density into a set of "natural bond orbitals"
(NBO) chosen in such a way that each orbital is occupied by,
ideally, two electrons \cite{Lowdin_PR55,Carpenter_TC88,Weinhold_CE01}.
We use the so-called Electrostatic potential scheme (ESP),
in which the partial charges 
are determined from the best fit of the actual electrostatic
potential of the molecule \cite{GaussianBook}. This method is not
only better justified physically for our purposes, but it is also less
dependent on the choice of basis set used in the calculations, see Table
\ref{partials}.

\begin{table}[htb]
\caption{\label{partials} Partial charges in coronene assigned to
carbon and hydrogen atoms according to the Mulliken \protect{\cite{Mulliken_JCP55}},
natural bond orbitals (NBO) \protect{\cite{Carpenter_TC88}}
and best fit electrostatic potential (ESP) \protect{\cite{GaussianBook}} schemes
(in units of elementary charge). C$_1$ are carbons from the
innermost chain (there are 6 of them), C$_2$ are carbons from the
intermediate carbon chain (6), C$_3$ are carbons from the outer
carbon chain (12). }
\hskip -0.4cm
\begin{tabular}{|c|c|c|c|c|c|c|c|c|c|c|c|c|}
\hline
 & \multicolumn{4}{|c|}{Mulliken} & \multicolumn{4}{|c|}{NBO}& \multicolumn{4}{|c|}{ESP} \\
\hline
        &\multicolumn{1}{|c|}{C$_1$}
        &\multicolumn{1}{|c|}{C$_2$}
        &\multicolumn{1}{|c|}{C$_3$}
        &\multicolumn{1}{|c|}{H}
        &\multicolumn{1}{|c|}{C$_1$}
        &\multicolumn{1}{|c|}{C$_2$}
        &\multicolumn{1}{|c|}{C$_3$}
        &\multicolumn{1}{|c|}{H}
        &\multicolumn{1}{|c|}{C$_1$}
        &\multicolumn{1}{|c|}{C$_2$}
        &\multicolumn{1}{|c|}{C$_3$}
        &\multicolumn{1}{|c|}{H}    \\
        \hline
 B3LYP/6-31+G(d) & \multicolumn{1}{|c|}{-0.346}
        &\multicolumn{1}{|c|}{0.741}
        &\multicolumn{1}{|c|}{-0.384}
        &\multicolumn{1}{|c|}{0.185}
        &\multicolumn{1}{|c|}{0.158}
        &\multicolumn{1}{|c|}{-0.138}
        &\multicolumn{1}{|c|}{-0.213}
        &\multicolumn{1}{|c|}{0.241}
        &\multicolumn{1}{|c|}{-0.002}
        &\multicolumn{1}{|c|}{0.129}
        &\multicolumn{1}{|c|}{-0.207}
        &\multicolumn{1}{|c|}{0.148} \\
        \hline
 B3LYP/D95 & \multicolumn{1}{|c|}{-0.008}
        &\multicolumn{1}{|c|}{0.463}
        &\multicolumn{1}{|c|}{-0.46}
        &\multicolumn{1}{|c|}{0.232}
        &\multicolumn{1}{|c|}{-0.016}
        &\multicolumn{1}{|c|}{-0.041}
        &\multicolumn{1}{|c|}{-0.19}
        &\multicolumn{1}{|c|}{0.218}
        &\multicolumn{1}{|c|}{0.002}
        &\multicolumn{1}{|c|}{0.13}
        &\multicolumn{1}{|c|}{-0.237}
        &\multicolumn{1}{|c|}{0.168} \\
        \hline
B3LYP/cc-pVDZ & \multicolumn{1}{|c|}{0.064}
        &\multicolumn{1}{|c|}{-0.02}
        &\multicolumn{1}{|c|}{0.026}
        &\multicolumn{1}{|c|}{-0.049}
        &\multicolumn{1}{|c|}{-0.008}
        &\multicolumn{1}{|c|}{-0.053}
        &\multicolumn{1}{|c|}{-0.199}
        &\multicolumn{1}{|c|}{0.229}
        &\multicolumn{1}{|c|}{-0.004}
        &\multicolumn{1}{|c|}{0.095}
        &\multicolumn{1}{|c|}{-0.171}
        &\multicolumn{1}{|c|}{0.129} \\
        \hline
PBE/6-31+G(d)& \multicolumn{1}{|c|}{-0.362}
        &\multicolumn{1}{|c|}{0.807}
        &\multicolumn{1}{|c|}{-0.42}
        &\multicolumn{1}{|c|}{0.197}
        &\multicolumn{1}{|c|}{0.115}
        &\multicolumn{1}{|c|}{-0.136}
        &\multicolumn{1}{|c|}{-0.22}
        &\multicolumn{1}{|c|}{0.251}
        &\multicolumn{1}{|c|}{-0.003}
        &\multicolumn{1}{|c|}{0.121}
        &\multicolumn{1}{|c|}{-0.223}
        &\multicolumn{1}{|c|}{0.148} \\
        \hline
PBE/D95 & \multicolumn{1}{|c|}{0.033}
        &\multicolumn{1}{|c|}{0.454}
        &\multicolumn{1}{|c|}{-0.482}
        &\multicolumn{1}{|c|}{0.238}
        &\multicolumn{1}{|c|}{-0.016}
        &\multicolumn{1}{|c|}{-0.044}
        &\multicolumn{1}{|c|}{-0.196}
        &\multicolumn{1}{|c|}{0.225}
        &\multicolumn{1}{|c|}{-0.003}
        &\multicolumn{1}{|c|}{0.133}
        &\multicolumn{1}{|c|}{-0.236}
        &\multicolumn{1}{|c|}{0.16} \\
        \hline
PBE/cc-pVDZ & \multicolumn{1}{|c|}{0.076}
        &\multicolumn{1}{|c|}{-0.014}
        &\multicolumn{1}{|c|}{0.009}
        &\multicolumn{1}{|c|}{-0.04}
        &\multicolumn{1}{|c|}{-0.008}
        &\multicolumn{1}{|c|}{-0.055}
        &\multicolumn{1}{|c|}{0.207}
        &\multicolumn{1}{|c|}{0.239}
        &\multicolumn{1}{|c|}{-0.0002}
        &\multicolumn{1}{|c|}{0.086}
        &\multicolumn{1}{|c|}{-0.18}
        &\multicolumn{1}{|c|}{0.13} \\
        \hline
PBE/cc-pVTZ & \multicolumn{1}{|c|}{0.371}
        &\multicolumn{1}{|c|}{-0.029}
        &\multicolumn{1}{|c|}{-0.335}
        &\multicolumn{1}{|c|}{0.164}
        &\multicolumn{1}{|c|}{-0.009}
        &\multicolumn{1}{|c|}{-0.054}
        &\multicolumn{1}{|c|}{-0.187}
        &\multicolumn{1}{|c|}{0.218}
        &\multicolumn{1}{|c|}{0.002}
        &\multicolumn{1}{|c|}{0.114}
        &\multicolumn{1}{|c|}{-0.205}
        &\multicolumn{1}{|c|}{0.149} \\
        \hline
\end{tabular}
\end{table}

In the simplest approximation, the electron density, shifted
from the hydrogens, is assigned to the outer carbons, covalently
bonded to the hydrogens.
Then the ring corresponding to the chain of the outermost carbons
is charged negatively.
The inner rings remain neutral.
The charge distribution in such approximation can be crudely described
as a polarized band (with a positively charged outer edge and
a negatively charged inner edge) located
along the perimeter of the molecule.
This charge distribution can also be considered as a dipole
continuously distributed along the outer border of the molecule.
In fact, within this approximation one can assign
a typical value of the partial charge (say, 0.15 \cite{vdWaal_JCP83}) to all the hydrogen atoms
and then one immediately obtains the charges of the two outer rings.
The DFT analysis of the electron density in this case is not necessary.
This approximation provides
a tool for studying interactions of large PAHs which cannot be
treated with \textit{ab initio} methods.

The interaction energy between two PAH molecules consists of a Coulomb
and a van der Waals terms calculated for each pair of the rings:
\begin{equation}
E = \sum_{\alpha,\beta} \left(E^{\rm \,Coul}_{\alpha \beta} +
E^{\rm \,vdW}_{\alpha \beta} \right),
\end{equation}
\noindent where index $\alpha$ enumerates the rings in one
monomer, and index $\beta$ enumerates the rings in the other monomer.

The energy of Coulomb interaction of two charged rings $E^{\rm \,Coul}_{\alpha \beta}$
can be found by integrating the electrostatic potential created
by one ring along the contour of the other ring:
\begin{equation}
E^{\rm \,Coul}_{\alpha \beta} = \frac{Q_{\beta}}{|\ell_\beta|}
\oint \phi^{\rm Coul}_{\alpha} (\bf r) \, {\rm d} \ell_{\beta},
\end{equation}
\noindent where $Q_\beta$ is the charge of the ring $\beta$ equal
to the sum of partial charges of all atoms belonging to the ring,
$|\ell_{\beta}|$ is the length of the ring.
The electrostatic potential $\phi^{\rm Coul}_{\alpha}(\bf r)$
created by a circular ring is expresses via the complete Elliptic function $K$,
\begin{equation}
\phi^{\rm Coul}_{\alpha} ({\bf r}) = \frac{2 Q_\alpha}{\pi} \frac{1}{\sqrt{2 r R_\alpha}}
\frac{1}{\sqrt{x+1}} \,
K\left(\frac{2}{x+1}\right),
\end{equation}
\begin{equation}
x=\frac{r^2+R_{\alpha}^{2}}{2 r R_{\alpha} \sin \theta}.
\end{equation}
\noindent Here $R_{\alpha}$ is the radius of the ring $\alpha$,
$r$ is the modulus of $\bf r$, $\theta$ is the angle between $\bf r$
and the ring normal.

The van der Waal's term $E^{\rm \,vdW}_{\alpha \beta}$ can be calculated
similarly,
\begin{equation}
E^{\rm \,vdW}_{\alpha \beta} = \oint
\left(
 \frac{Q_{\beta}^{\rm W}}{|\ell_\beta|} \phi^{\rm W}_{\alpha} ({\bf r}) +
 \frac{Q_{\beta}^{\rm X}}{|\ell_\beta|} \phi^{\rm X}_{\alpha} ({\bf r})
\right)
 \, {\rm d} \ell_{\beta}.
\end{equation}
\noindent The van der Waals charges $Q^{\rm W}$ and $Q^{\rm X}$
are defined as sums of the $C_6$ and $C_{12}$ coefficients, respectively,
of all the atoms belonging to the ring. For a circular ring
the van der Waals potentials $\phi^{\rm W}$ and $\phi^{\rm X}$
are expressed via elementary functions,
\begin{equation}
\phi^{\rm W}_{\alpha} ({\bf r}) = \frac{Q_\alpha^{\rm W}}{2} \frac{1}{(2 r R_\alpha)^3}
\frac{2x^2+1}{(x^2-1)^{5/2}},
\end{equation}
\begin{equation}
\phi^{\rm X}_{\alpha} ({\bf r}) = \frac{Q_\alpha^{\rm X}}{8} \frac{1}{(2 r R_\alpha)^6}
\frac{x(8x^4+40x^2+15)}{(x^2-1)^{11/2}}.
\end{equation}

The model predicts the T-shaped and PD configurations to be stable,
in accordance with the results of the DFT calculations.
In addition to it, the model provides an intutively obvious
explanation why even for such a large PAH
as coronene the T-shaped configuration
is comparable in energy with the stack configuration.
This happens due to the round shape of the coronene molecule,
since only the "edge to plane" configurations can be stable
electrostatically.
On the other hand, the dispersive interaction favors superimposed stacks
for which the "contact area" is largest.
For the non-round molecules, e.g., anthracene,
the interplay between these two factors
results in parallel-displaced or rotated configurations \cite{Lim03},
because a parallel displacement or a rotation can significantly
decrease the electrostatic repulsion while not seriosly affecting the
van der Waals attraction.
However, for the round molecules, such as coronene,
the rotation or parallel displacement are much less
effective in decreasing the Coulomb repulsion
and the T-shaped configuration can become competitive.

Therefore, we conclude that
in the dimerization process (and in formation of larger clusters of PAH
molecules) the shapes of the molecules are very important.
The proposed  simple model
can be useful in
qualitative analysis of the possible configurations and
in making the results of such analysis intuitively clear.
The model can be applied to any
polycyclic aromatic hydrocarbons, and it
can be easily generalized on the DNA bases, etc.
The model can also be used for predicting structures of
larger coronene clusters and clusters of other PAHs.

\acknowledgements This work is partially supported by the European
Commission within the Network of Excellence project EXCELL, by INTAS
under the grant 03-51-6170 and by the Russian Foundation for Basic
Research under the grant 06-02-17227-a. We are grateful to Dr. A.
Korol for fruitful discussions and Dipl. Phys. I. A. Solov'yov for
providing us with a useful graphical software for drawing
molecular images.
We acknowledge access to the computer cluster at
the Center for Scientific Computing of the Johann Wolfgang
Goethe-University where the computations have been performed.

\end{document}